# Open data to evaluate academic researchers: an experiment with the Italian Scientific Habilitation

Angelo Di Iorio[1] and Silvio Peroni[3] and Francesco Poggi[2]

*[1] angelo.diorio@unibo.it - [2] francesco.poggi @unibo.it*
Department of Computer Science, University of Bologna, Bologna (Italy)

*[3] silvio.peroni@unibo.it*
Department of Classical Philology and Italian Studies, University of Bologna, Bologna (Italy)

**Abstract**

The need for scholarly open data is ever increasing. While there are large repositories of open access articles and free publication indexes, there are still a few examples of free citation networks and their coverage is partial. One of the results is that most of the evaluation processes based on citation counts rely on commercial citation databases. Things are changing under the pressure of the Initiative for Open Citations (I4OC), whose goal is to campaign for scholarly publishers to make their citations as totally open. This paper investigates the growth of open citations with an experiment on the Italian Scientific Habilitation, the National process for University Professor qualification which instead uses data from commercial indexes. We simulated the procedure by only using open data and explored similarities and differences with the official results. The outcomes of the experiment show that the amount of open citation data currently available is not yet enough for obtaining similar results.

## Introduction

Citations indexes are becoming more and more important for evaluating scientific performances of a given research body. For instance, in many countries citation and bibliometric indicators are one of the factors that can be used for assessing individuals or institutions to allocate funding at national level: in Germany the impact factor of the publications is used in performance-based funding systems, in Finland the reallocation system uses bibliometrics and citation indexes as ones of the considered measures, in Norway a two-level bibliometric indicator is used for similar purposes, etc. (Vieira et al., 2014a).

Several works analyzed the relation between citation indexes and research assessment procedures. At national level, the relation between bibliometric indicators and the results of the Research Assessment Exercise (RAE) in Britain (Norris and Oppenheim, 2003; Taylor, 2011) or the Italian Triennial Assessment Exercise (VTR) (Abramo et al., 2009; Franceschet and Costantini, 2011) have been investigated. Other studies focused on the assessments of departments (Aksnes, 2003) and research groups (Van Raan, 2006). Just a few works have been made at the individual level (Nederhof and Van Raan, 1987; Bornmann and Daniel, 2006; Bornmann et al., 2008), while many analyzed the correlation between indicators and research performances (Leydesdorff, 2009; Franceschet, 2009). Recent works analyzed the correlation between traditional bibliometric indicators and altmetrics by also taking into account quality assessment procedures performed by peers (Nuzzolese et al., 2018; Wouters et al., 2015; Bornmann and Haunschild, 2018).

In this work we focus on the analysis of the Italian National Scientific Habilitation (ASN), a nation-wide evaluation process introduced some years ago by Law 240/2010 (Law dec. 30, n. 240, 2011) for University Professor position recruitment. The ASN is similar to other habilitation procedures already in place in other countries in that it is a prerequisite for becoming a university professor.

The procedure is based on scientific qualification criteria that take into account, among other factors, bibliometric indicators such as the number of citations and the h-index of the candidates. Citation data are taken from commercial databases, as it happens in other countries. One of the reasons is that the open citation indexes are still a few and their coverage is limited (van Eck et al., 2018). This is an issue not only for evaluation procedures but also for research activities on open science, trends and topics analysis, scientometrics and so on. The 'Initiative for Open Citations' (I4OC, https://i4oc.org) has been launched to gather publishers, researchers, and other interested parties and to promote the "unrestricted availability of scholarly citation data". The movement is gaining momentum and making available a lot of free citation data.

This work is part of a larger effort, whose goal is to monitor the growth of these data and their relation with closed ones. We are not only interested in counting open citations but also in exploring their distribution among datasets and domains, and their applicability to evaluation tasks. To the best of our knowledge, this is the first attempt to look at open citations for these tasks.

Here we would like to answer the following research questions:

RQ1.    To what extent open bibliographic metadata and open citation data can be used for evaluation purposes today?

RQ2.    Which open data would produce results comparable to those of closed ones?

RQ3.    Is there any case in which a negative evaluation would turn into a positive one, if open data were used instead of closed ones?

To answer these questions we run an experiment on the Italian Habilitation in the Computer Science domain. The test gave us valuable indications and allowed us to build the overall infrastructure for extending our analysis to other domains.

One of the reasons for starting with Computer Science was the availability of open, complete and well-maintained repositories of articles and publication lists. In fact, our experiment consisted of two phases:

1.  computing the indicators proposed by the ASN for all candidates by only taking into account open data. We collected these data from three main sources, namely Crossref (https://www.crossref.org/), DBLP (https://dblp.uni-trier.de/) and COCI (http://opencitations.net/index/coci) that will be introduced in the following sections;

2.  comparing the outcome of such evaluation with the official one, whose data were collected from Scopus and Web of Science.

The experiments showed that there is still a quite large gap between open and closed citations and the former cannot yet be used directly for these tasks (RQ1). However, the data about the types of the publications, in particular journals, are comparable with the outcomes of the ANS 2016 (RQ2). Interestingly, we also found a few candidates for which open data would change the evaluation from negative to positive (RQ3).

The paper presents the methods and the results of our experiment and its implications. It is then structured as follows: Section "Background" provides some background introducing both the ASN process and the open citations status. Section "Methods and materials" introduces the sources we used for gathering the metadata and citation data, while Section "Experiments with open data and ASN" explains our experiment in detail. Results and lessons learned are discussed in Section "Results", before concluding and drafting new research directions in Section "Discussion and conclusions".

**Background**

In order to introduce our experiment we first need to provide readers with some background about the Italian ASN and the I4OC movement.

*The Italian National Scientific Habilitation (ASN)*

The ASN (Law dec. 30, n. 240, 2011) is a nation-wide research assessment procedure similar to others already in place in other countries. The first two sessions of the ASN took place in 2012 and 2013, followed by other sessions in 2016, 2017 and 2018. The ASN is meant to attest that an individual has reached the scientific maturity required for applying for a specific role as Associate or Full Professor. Each candidate is bound to a specific Recruitment Field (RF), which corresponds to a scientific field of study. RFs are organized in groups, which are in turn sorted in 14 Scientific Areas (SAs). The assessment of the candidates of each discipline (RF) is performed by a committee of full professors, which evaluates the CVs submitted by the applicants. The evaluation also takes into account *three quantitative indicators* computed for each candidate.

The ASN introduced two types of indicators: *bibliometric* and *non-bibliometric*. Bibliometric indicators apply to scientific disciplines for which reliable citation databases exist, among which Computer Science, on which we performed our analysis. The three bibliometric indicators are:

A. Normalized number of journal papers;
B. Normalized number of citations received;
C. Normalized h-index.

Since citations and paper count increase over time, normalization based on the scientific age (the number of years since the first publication) is used to compute these indicators more reliably.

The three indicators are computed by ANVUR – the National Agency in charge of the Habilitation process – for each candidate, starting from the data in Scopus and Web of Science. These databases, in fact, contain either the full list of classified publications for each candidate (used to compute indicator A) and the full list of citations received by each article (used to compute indicators B and C). These data were automatically compiled into a CV in PDF, submitted to the committee.

The preliminary step of the ASN consisted of checking, for each candidate, how many indicators exceeded some thresholds. The candidates were required to exceed at least two indicators over three. Exceeding thresholds does not imply that the candidate gets the habilitation automatically but is only an indication for the committee.

Though, this step is the focus of our experiment. We do not analyze the final subjective evaluation of the committee but we compare the ability of each candidate to exceed thresholds when using open or closed data.

The thresholds were computed by ANVUR as well and officially released for each RF. Even in this case, data to compute thresholds were taken from Scopus and Web of Science. In particular, in 2012, the thresholds were defined as the medians of the values computed for all Associate and Full professors already permanent. However, in 2016, the values were established by ANVUR but they did not disclose the algorithm to do that.

Several analyses of the ASN process and results have been carried out by the research community, like the quantitative analyses of ASN 2012 in (Marzolla, 2015) and (Marzolla, 2016), the study on the impact of the ASN on self-citation rate (Scarpa et al., 2018), the analysis on the relationship of the ASN outcomes to the actual scientific merit of candidates (Abramo and D'Angelo, 2015), etc.. Our goal is not to evaluate the reliability of ASN, nor to assess its effects and consequences, but to investigate to what extent such an evaluation could be performed without using commercial citation indexes.

*The open citations movement*

The first project to introduce for the very first time the open availability of open bibliographic and citation data by the use of Semantic Web (Linked Data) technologies was the

OpenCitations Corpus, in 2010, as the main output of a project funded by JISC (Shotton, 2013). However, the availability of open citation data recently changed drastically with the introduction of Initiative for Open Citations (I4OC, https://i4oc.org), in April 2017.

The Initiative was born with the idea of promoting the release of open citation data, and explicitly asked the main scholarly publishers, who deposited their citations on Crossref (https://crossref.org), to release them in the public domain. As a result, now we have several millions of citation data openly available on the Web, a list of important stakeholders – such as libraries, consortiums, projects, organizations, companies, and, in particular, founders (Shotton, 2018) – supporting the movement, several international events (e.g. the Workshop on Open Citations and WikiCite 2018) organised for promoting the open availability of citation data, and several projects and datasets have been released so far so as to leverage the open citation data available online. As a result, there is a growing list of publishers that release their citation data in Crossref, and these also includes citation data of important Computer Science venues and publishers such as the Association for Computing Machinery (ACM).

**Methods and materials**

The first step of our analysis consisted in computing the indicators proposed by the ASN for all the candidates in the Recruitment Field *Informatics* by taking into account only open data. To do so, we first collected the curricula of all applicants to the five sessions of the ASN 2016, which have been made publicly available on the ANVUR website for a short period of time. We collected 518 CVs for level I (full professor) and 757 CVs for level II (associate professor). Note that each CV correspond to a single application, and that the same applicant may apply multiple times (i.e. in more than one session) for multiple levels.

The next step consisted in collecting the list of the DOIs of all the publications that each candidate specified in her/his CV, thus excluding all the publications that do not have associated a DOI. This lead us to miss some publications, for instance the workshop articles in the CEUR-WS volumes which are published without a DOI (though Scopus takes track of them). However, we expect that the loss in term of citations is rather limited, considering that the most relevant works and their extensions usually go to journal articles and conference proceedings papers, which are instead associated with a DOI.

We used two different sources to retrieve the features needed for such computation:

1. DBLP (https://dblp.uni-trier.de): it is a free and publicly available computer science bibliography repository started in 1993 at the University of Trier, Germany. DBLP contained more than 4.4 million bibliographic entries (as of January 2019). We search the candidates by name using the DBLP API, and downloaded the full publication list of each of them. We exploited standard disambiguation techniques and ORCID data to identify candidates.

2. Candidates' CVs: we extracted the text from each CV (which was originally in PDF format), and searched for valid DOIs using a simple pattern matching approach to produce the publication list. The DOI system Proxy Server REST API (http://www.doi.org/factsheets/DOIProxy.html#rest-api) have been used to verify the existence and validity of the collected DOIs.

The collected data have been used to produce three publication lists for each candidate: the first contains the DOIs of the publications retrieved from DBLP, the second contains the DOIs of the publications extracted from the CV, and the latter is the union of the DOI of the publications collected from the two sources (where duplicates DOIs have been considered only once). Table 1 reports some basic statistics about the dataset generated as output of this step.

| Level | CVs | DOI DBLP | DOI CV | DOI UNION |
|---|---|---|---|---|
| | | *(\* in parentheses average DOIs per applicant)* | | |
| *Associate Professor* | 757 | 31713 (41.9) | 31896 (42.1) | 36820 (48.6) |
| *Full Professor* | 518 | 37728 (72.8) | 37793 (73.0) | 42375 (81.8) |

**Table 1. Basic statistics about the application submitted for Recruitment Field *Informatics* at the ASN 2016. For each level we report the number of CVs collected, and the overall number of retrieved DOIs of applicants' publications and, in parentheses, the average number of publications per candidate extracted from (i.) DBLP, (ii) the CVs, and (iii) the union of them. In addition, both DBLP and Crossref were used to retrieve the publication types of all the aforementioned DOIs.**

All the citations related to the DOIs extracted were gathered from COCI, the OpenCitations Index of Crossref open DOI-to-DOI citations (http://opencitations.net/index/coci). This dataset is provided by OpenCitations, a scholarly infrastructure organization dedicated to open scholarship and the publication of open bibliographic and citation data (Peroni and Shotton, 2018) by the use of Semantic Web (Linked Data) technologies. Launched in July 2018, COCI is the first of the Indexes proposed by OpenCitations (http://opencitations.net/index) in which citations are exposed as first-class data entities with accompanying properties, and currently contains 449,840,503 DOI-to-DOI citation links between 46,534,705 distinct bibliographic entities (OpenCitations, 2018).

To date, the majority of the citations stored in Crossref that are not available in COCI comes from just three publishers: Elsevier, the American Chemical Society and University of Chicago Press (Heibi et al., 2019). This is due to the particular access policy chosen by these publishers, since such citation data refer to publications for which the reference lists are not visible to anyone outside the Crossref Cited-by membership. The advantage to access COCI instead of Crossref is that COCI also contains DOI-to-DOI citations that are included in the Crossref 'limited' dataset, which is accessible only to users of the Crossref Cited-by service and to Metadata Plus members of Crossref. Additional information about the way Crossref classifies reference lists is available at https://www.crossref.org/reference-distribution/. The fact that COCI does not contain citations from Elsevier's articles can be a bottleneck to the study we are presenting, since such publisher manages several of the most important Computer Science journals that are valuable sources of citations to other Computer Science articles.

The source code of the pipeline to collect these data is available as open source at https://github.com/sosgang/asn2016-issi2019, while the data are available on Zenodo and are released with a CC0 waiver (Di Iorio, Peroni and Poggi, 2019).

## Experiments with open data and ASN

The core of the experiment consisted of studying the performance of the candidates when using open data instead of closed ones. We repeated the test under three conditions corresponding to the three overlapping sets of DOIs. For the sake of clarity these conditions will be indicated with an acronym from now on:

- **CCV**: DOIs taken from CVs and citations taken from COCI;
- **CDBLP**: DOIs taken from DBLP and citations taken from COCI;
- **CU**: DOIs obtained as the union of the DOIs in CCV and CDLP, and citations taken from COCI.

The following step consists of calculating the three thresholds against which compare our data. Our initial plan was to re-calculate these thresholds as medians of the values of the indicators for Associate and Full Professors, as done for the ASN 2012. We also expected to compute the indicators for each condition. This was not done since ANVUR did not publish the algorithm to calculate the official thresholds in 2016 but only their values. Then we used the official thresholds directly, even if they were calculated from closed (and, potentially, more rich) data. This is not optimal but gave us valuable insights and we plan to do further experiments on the thresholds as discussed at the end of the paper. The current values are shown in Table 2.

**Table 2. The ASN 2016 thresholds for the Associated Professor and Full Professor positions.**

| Role | #journals articles (A) | #citations (B) | H-index (C) |
|------|------------------------|----------------|-------------|
| Associate Professor | 8 | 216 | 8 |
| Full Professor | 5 | 118 | 6 |

Then, for each indicator we calculated the percentage of candidates who were able to exceed the thresholds in both our test and the official ASN. We also measured the amount of candidates who exceeded two thresholds over three - and thus were able to continue the process to get the qualification - in both cases. Note that we do not compare the values of the indicators directly, as we expected them to be different, rather their contribution to the habilitation.

**Table 3. Two (real) candidates of the ASN accompanied by their values for the three indicators used in the ASN, i.e. number of journal articles, number of citations, and h-index. The number shown refers to those ones retrieved by means of open data and the real ones calculated in the context of the ASN.**

| id | Open data | | | Official ASN data | | |
|----|-----------|---|---|-------------------|---|---|
| | #journal articles (A) | #citations (B) | H-index (C) | #journals articles (A) | #citations (B) | H-index (C) |
| 1 | 15 | 417 | 12 | 17 | 1144 | 17 |
| 2 | 8 | 197 | 7 | 33 | 1939 | 18 |

For instance, let us consider the two (real) candidates in Table 3. They both applied for the qualification as Full Professor and exceeded all three thresholds. The values of their indicators were lower when we only took open data into account in both cases. However, candidate #1 was able to exceed two thresholds anyway. The same did not happen for candidate #2. We counted the percentage of these situations to study the relation between open and closed data. The measurements were also repeated on the three datasets CCV, CDBLP and CU in order to get a more precise picture and are fully described in the next section.

**Results**

We measured the percentage of candidates, for all levels and under all conditions, for which there is agreement between our test and the official ASN outcome. Table 4 summarizes the data on the 517 candidates as Full Professor (Level 1). The three columns correspond to the three conditions introduced CCV, CDBLP and CU. The rows detail each indicator and the overall result (two indicators over three above/below the thresholds).

**Table 4. The percentage of candidates as Full Professor who achieved the same result in our open data simulation and the official ASN, for each indicator and under each condition.**

| Full Professor (518 candidates) | | | |
|---|---|---|---|
| | CCV | CDBLP | CU |
| *Overall* | 59.07% | 58.88% | 67.95% |
| *Journals* | 89.77% | 89.58% | 93.82% |
| *Citations* | 50.77% | 50.58% | 58.49% |
| *h-index* | 59.65% | 59.46% | 67.76% |

Overall, the results on open data are not yet comparable to those on closed ones. In fact, the agreement ranges from 58.88% of CDBLP to 67.95% of CU. The three indicators contributed in different ways to this result: while there was a substantial agreement on indicator A (articles in journals) with a percentage of about 90% for all three cases, the percentages lower to about 50% for the citations (indicator B) and 60% for the h-index (indicator C). It is also worth noticing that the results increase of about 8-9% when considering the union of CCV and CDBLP.

Table 5 shows the results for the applications as Associate Professor. The number of candidates was 757 and the overall agreement was in line with the previous scenario. In fact about 57% of the candidates got the same result in both the evaluations for CCV and DBLP while the agreement grows up to 70.94% for CU. Again, the agreement was very high for the indicator A (journal articles) and the ratio between the three indicators was quite stable.

**Table 5. The percentage of candidates as Associate Professor who achieved the same result in our open data simulation and the official ASN, for each indicator and under each condition.**

| Associate Professor (757 candidates) | | | |
|---|---|---|---|
| | CCV | CDBLP | CU |
| *Overall* | 57.60% | 56.80% | 70.94% |
| *Journals* | 80.58% | 79.79% | 90.36% |
| *Citations* | 49.14% | 48.75% | 60.24% |
| *h-index* | 62.35% | 61.56% | 73.98% |

As expected, the overall trend was that candidates get worse results when only considering open citation data, since the amount of these data was still limited when compared with closed citation data. We also asked ourselves if there are instead candidates whose results improved if the ASN had used open data. It might also happen in fact that DBLP data (i.e. the DOIs it contains) are richer than the corresponding in Scopus and Web of Science, so that some indicators could differ.

To study such aspect we measured the percentage of candidates that exceeded the thresholds with open data but not with the closed ones (and vice versa). We also computed these variations for all indicators and the overall score. Results are summarized in Table 6, under the three conditions CCV, CDBLP and CU.

**Table 6. The percentage of candidates who exceeded the thresholds with open data but not with the closed ones (column '+') or vice versa (column '-'). The results are shown for all conditions CCV, CDBLP and CU. The table is split in two mirror-like parts, for candidates as Full and Associate Professor.**

| | CCV | | CDBLP | | CU | |
|---|---|---|---|---|---|---|
| | + | - | + | - | + | - |
| *Full Professor* | | | | | | |
| *overall* | 0.19% | 40.73% | 0.19% | 40.93% | 0.39% | 31.66% |
| *journals* | 1.54% | 8.69% | 1.54% | 8.88% | 2.32% | 3.86% |
| *citations* | 0.00% | 49.23% | 0.00% | 49.42% | 0.39% | 41.12% |
| *h-index* | 0.19% | 40.15% | 0.19% | 40.35% | 0.19% | 32.05% |
| *Associate Professor* | | | | | | |
| *overall* | 0.13% | 42.27% | 0.13% | 43.06% | 0.13% | 28.93% |
| *journals* | 2.77% | 16.64% | 2.77% | 17.44% | 3.96% | 5.68% |
| *citations* | 0.26% | 50.59% | 0.13% | 51.12% | 0.53% | 39.23% |
| *h-index* | 0.66% | 36.99% | 0.66% | 37.78% | 1.06% | 24.97% |

In a very limited set of cases the open data produced a growth in the performance, with a slightly more evident increment for the indicator A. In general the impact of adopting open data would then be very limited with a few exceptions.

**Discussion and conclusions**

The results of our experiment are in line with what we expected with some interesting unexpected behavior. First of all, it is evident that **open citation data are not yet complete to substitute the closed data** used by ANVUR within the ASN in Computer Science. This answer our research question RQ1. It was foreseen considering that several publishers have not released their citation data as open and there is still a gap between the two sets and **some effort is still needed to convince publishers to release their data**. On the other hand, the overall agreement of around 60% is a positive result that make us optimistic about the possibility of performing some reliable evaluation on these data as well.

The high agreement on the indicator A (journal articles) allowed us to answer the research question RQ2. The classification of the publications is extremely good in open data and we speculate that the fact that the agreement is not full is due to the lack of data instead of their inaccuracy. The **accuracy of open and closed data on the classification of the publication venues (in particular "journals" vs. "non-journals") is comparable**, and counting journals of publication by type can be done reliably.

While the indicator A proved to be stable we witnessed a remarkable improvement of the agreement in the CU scenario compared to the CDBLP or CCV ones. The overall agreement, for instance, goes from to 56% to 70%, and from 63% to 70% on the h-index. We speculate that is a consequence of the nature and the coverage of the two datasets, in relation of the scientific production of the candidates. The DBLP source in fact is very accurate for pure computer scientists since almost all their publications are listed there. On the other hand, DBLP misses articles in close domains, such as bioinformatics and physics. There were candidates that applied for the habilitation in Computer Science even if they are experts in these domains. We are not interested here in the overlap between the two domains, neither on the evolution of the research topics, studied for instance in Osborne et al. (2013) and Salatino et al. (2017), but we noticed that some candidates were penalized by the mono-disciplinary approach of the CDBLP scenario. The CCV condition, on the other hand, produced slightly better results for multidisciplinary experts, with some penalisation for pure computer scientists. The union of these two inevitably had a positive impact on the agreement, that raised up to 70%. Our conclusion is that **combining different sources of open data is a fundamental step to better evaluate candidates**.

A further conclusion that we have drawn from our data is that there are no substantial differences between the simulation on Full and Associate Professors. The overall distribution of percentages and relation between indicators is basically the same, even if specific data are different. Contrarily to other aspects of the ASN in which different behaviors of candidates were pointed out (Marini, 2017) our experiment showed that an evaluation based on **open data works in the same way for either Full or Associate Professors assessment**.

One objection that could be raised on our work is that we used the ASN official thresholds even on open data, that we already knew were less. We are aware that this is not optimal but we had no undisputable algorithm to re-calculate thresholds on our dataset in a consistent way with the ASN procedure, since details about that were not yet published by ANVUR. As discussed earlier, in fact, the ASN thresholds were computed automatically in 2012 and made available by law in the following sessions without further details.

To study thresholds, we artificially tuned them for open data and looked at how the agreement on the candidates, overall and for each indicator, changed when changing these values. Note that we only manipulated our simulated thresholds leaving untouched the official ones. The variation is plotted in the following diagram for the candidates as Associate Professors in the CDBLP condition. For the sake of brevity we only show this scenario but there are no significant differences with the other ones.

**Figure 1. The variation of the agreement between our simulation and the official ASN, when lowering the thresholds for open data. The X axis indicates the ratio between the new simulated thresholds and the original ones; the Y axis indicates the amount of candidates on which there is agreement. Data are shown for candidates as Associate Professor under the DBLP condition.**

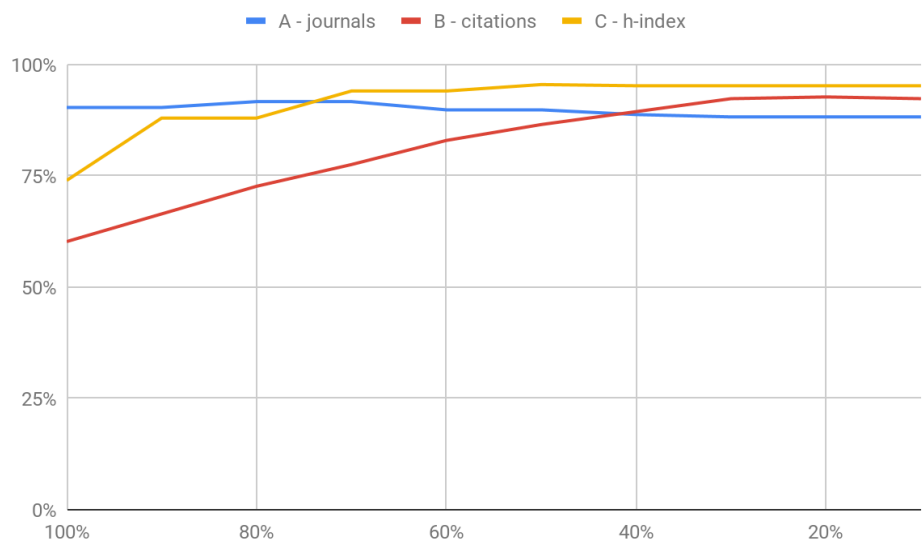

The X axis indicates the ratio between the new simulated thresholds and the official ones; the Y axis indicates the amount of candidates on which there is agreement; the three lines show the behavior of the three indicators. Thus, the value labelled as 100% on the X axis correspond to the official ASN thresholds; the position 60% correspond to the values 3

(indicator A), 71 (indicator B) and 4 (indicator C) calculated in percentage to the original values 5 (indicator A), 118 (indicator B) and 6 (indicator C).

It is interesting to notice that the indicator A is very stable. This is in line with previous results since there was already high agreement on this indicator and thus the effect of lowering the threshold is limited. Note also that a lower threshold might also reduce the agreement since a candidate might exceed the threshold in the simulated ASN but not on the official one. This is the reason why the indicator goes slightly down.

It is also evident a growth of the agreement when reducing the other two thresholds by 20-30%. This mitigates the limited availability of open data since candidates are given the possibility to exceed thresholds anyway. Note that the agreement never goes up to 100% since lower thresholds might change the performance of some candidates; this also depends on the fact that here we are considering each indicator separately while the agreement is computed on two of them.

In the near future we plan to also study the interaction between these indicators and to put in place optimization techniques to investigate thresholds for open data. Such an analysis in fact is limited since all thresholds are lowered proportionally, while our work highlighted clear differences between indicators caused by the differences between the data sources.

Another activity that we plan for the future is to extend our analysis research to other domains and to dig into the potentialities and limitations of other open repositories, with particularly attention to the field of medicine and open access journals.

# References

Abramo, G., & D'Angelo, C. A. (2015). An assessment of the first "scientific habilitation" for university appointments in Italy. Economia Politica, 32(3), 329-357.

Abramo, G., D'Angelo, C. A., and Caprasecca, A. (2009). Allocative efficiency in public research funding: Can bibliometrics help? Research Policy, 38(1):206–215. DOI:10.1016/j.respol.2008.11.001.

Aksnes, D. (2003). A macro study of self-citation. Scientometrics, 56(2):235–246. DOI:10.1023/A:102191922.

Bauin, S. & Rothman, H. (1992). "Impact" of journals as proxies for citation counts. In P. Weingart, R. Sehringer & M. Winterhager (Eds.), *Representations of Science and Technology* (pp. 225-239). Leiden: DSWO Press.

Borgman, C.L. (Ed.). (1990). *Scholarly Communication and Bibliometrics*. London: Sage.

Bornmann, L. and Daniel, H.-D. (2006). Selecting scientific excellence through committee peer review-A citation analysis of publications previously published to approval or rejection of post-doctoral research fellowship applicants. Scientometrics, 68(3):427–440. DOI:10.1007/s11192-006-0121-1.

Bornmann, L. and Haunschild, R. (2018). Do altmetrics correlate with the quality of papers? A large-scale empirical study based on F1000Prime data. PloS one, 13(5):e0197133. DOI:10.1371/journal.pone.0197133.

Bornmann, L., Wallon, G., and Ledin, A. (2008). Does the committee peer review select the best applicants for funding? An investigation of the selection process for two European molecular biology organization programmes. PLoS One, 3(10):e3480. DOI:10.1371/journal.pone.0003480.

Buckland, M. & Gey, F. (1994). The relationship between recall and precision. *Journal of the American Society for Information Science*, 45, 12-19.

Di Iorio, A., Peroni, S., Poggi, F. (2019). ASN 2016 evaluation with open data. Version 1. Zenodo. DOI: 10.5281/zenodo.2559481

Franceschet, M. and Costantini, A. (2011). The first Italian research assessment exercise: A bibliometric perspective. Journal of informetrics, 5(2):275–291. DOI:10.1016/j.joi.2010.12.002.

Heibi, I., Peroni, S. & Shotton, D. (2019). Crowdsourcing open citations with CROCI – An analysis of the current status of open citations, and a proposal. Submitted for publication at the 17th International Conference on Scientometrics and Bibliometrics (ISSI 2019). Preprint available at

https://opencitations.wordpress.com/2019/02/07/crowdsourcing-open-citations-with-croci/ (last visited 7 February 2019)

Hoppe, K., Ammersbach, K., Lutes-Schaab, B. & Zinssmeister, G. (1990). EXPRESS: An experimental interface for factual information retrieval. In J.-L. Vidick (Ed.), *Proceedings of the 13th International Conference on Research and Development in Information Retrieval* (ACM SIGIR '91) (pp. 63-81). Brussels: ACM.

Kling, R. & Elliott, M. (1994). *Digital library design for usability*. Retrieved December 7, 2001 from: http://www.csdl.tamu.edu/DL94/paper/kling.html.

Law dec. 30, n. 240 (2011). Rules concerning the organization of the universities, academic employees and recruitment procedures, empowering the government to foster the quality and efficiency of the university system (Norme in materia di organizzazione delle università, di personale accademico e reclutamento, nonche' delega al Governo per incentivare la qualità e l'efficienza del sistema universitario), Gazzetta Ufficiale n. 10 del 14 gennaio 2011 - Suppl. Ordinario n. 11. Available at http://www.gazzettaufficiale.it/eli/id/2011/01/14/011G0009/sg. (Accessed 9 October 2018).

Marini, G., 2017. New promotion patterns in Italian universities: Less seniority and more productivity? Data from ASN. Higher Education, 73(2), pp.189-205.

Marzolla, M. (2015). Quantitative analysis of the Italian national scientific qualification. Journal of Informetrics, 9(2):285–316. DOI:10.1016/j.joi.2015.02.006

Marzolla, M. (2016). Assessing evaluation procedures for individual researchers: The case of the Italian National Scientific Qualification. Journal of Informetrics, 10(2), 408-438. DOI: 10.1016/j.joi.2016.01.009

Nederhof, A. J. and Van Raan, A. F. (1987). Peer review and bibliometric indicators of scientific performance: a comparison of cum laude doctorates with ordinary doctorates in physics. Scientometrics, 11(5-6):333–350. DOI:10.1007/BF02279353.

Norris, M. and Oppenheim, C. (2003). Citation counts and the Research Assessment Exercise V: Archaeology and the 2001 RAE. Journal of Documentation, 59(6):709–730. DOI:10.1108/00220410310698734

Nuzzolese, A. G., Ciancarini, P., Gangemi, A., Peroni, S., Poggi, F., & Presutti, V. (2018). Do altmetrics work for assessing research quality?. Scientometrics, 1-24. DOI:10.1007/s11192-018-2988-z

OpenCitations (2018). COCI CSV dataset of all the citation data. Version 3. Figshare. DOI: 10.6084/m9.figshare.6741422.v3

Osborne, Francesco, Enrico Motta, and Paul Mulholland. "Exploring scholarly data with rexplore." In International semantic web conference, pp. 460-477. Springer, Berlin, Heidelberg, 2013.

Peroni, S. & Shotton, D. (2018). Open Citation: Definition. Version 1. Figshare. DOI: 10.6084/m9.figshare.6683855

Salatino, A.A., Osborne, F. and Motta, E., 2017. How are topics born? Understanding the research dynamics preceding the emergence of new areas. PeerJ Computer Science, 3, p.e119.

Scarpa, F., Bianco, V., & Tagliafico, L. A. (2018). The impact of the national assessment exercises on self-citation rate and publication venue: an empirical investigation on the engineering academic sector in Italy. Scientometrics, 117(2), 997-1022.

Shotton, D. (2013). Publishing: Open citations. Nature 502: 295–297. DOI: https://doi.org/10.1038/502295a

Shotton, D. (2018). Funders should mandate open citations. Nature, 553(7687), 129-129.

Taylor, J. (2011). The assessment of research quality in UK universities: peer review or metrics? British Journal of Management, 22(2):202–217. DOI:10.1111/j.1467-8551.2010.00722.x.

van Eck, N.J., Waltman, L., Larivière, V. & Sugimoto, C. R. (2018). Crossref as a new source of citation data: A comparison with Web of Science and Scopus. CWTS Blog. https://www.cwts.nl/blog?article=n-r2s234 (last visited 26 January 2018)

Van Raan, A. F. (2006). Comparison of the hirsch-index with standard bibliometric indicators and with peer judgment for 147 chemistry research groups. Scientometrics, 67(3):491–502. DOI:10.1556/Scient.67.2006.3.10.

Vieira, E. S., Cabral, J. A., and Gomes, J. A. (2014a). Definition of a model based on bibliometric indicators for assessing applicants to academic positions. Journal of the Association for Information Science and Technology, 65(3):560–577. DOI:10.1002/asi.22981.

Wouters, P., Thelwall, M., Kousha, K., Waltman, L., de Rijcke, S., Rushforth, A., and Franssen, T. (2015). The metric tide: Correlation analysis of REF2014 scores and metrics (Supplementary Report II to the Independent Review of the Role of Metrics in Research Assessment and Management). London: Higher Education Funding Council for England (HEFCE). DOI:10.13140/RG.2.1.3362.4162.